\documentclass[12pt]{article}

\begin{document}

\baselineskip=24pt plus 1pt minus 1pt

\begin{center}

{\large \bf Connecting the X(5)-$\beta^2$, X(5)-$\beta^4$, and
X(3) models to the shape/phase transition region of the
interacting boson model}

\bigskip\bigskip

{E. A. McCutchan$^{a}$\footnote{e-mail:
elizabeth.ricard-mccutchan@yale.edu}, Dennis
Bonatsos$^{b}$\footnote{e-mail: bonat@inp.demokritos.gr}, N. V.
Zamfir$^{c}$\footnote{e-mail: zamfir@tandem.nipne.ro} }

\bigskip

{$^{a}$ Wright Nuclear Structure Laboratory, Yale University,}

{New Haven, Connecticut 06520-8124, USA}

{$^{b}$ Institute of Nuclear Physics, National Center for
Scientific Research ``Demokritos'',}

{GR-15310 Aghia Paraskevi, Attiki, Greece}

{$^{c}$ National Institute of Physics and Nuclear Engineering,
Bucharest-Magurele, Romania}

\end{center}

\bigskip\bigskip
\centerline{\bf Abstract} \medskip

\medskip
The parameter independent (up to overall scale factors)
predictions of the X(5)-$\beta^2$, X(5)-$\beta^4$, and X(3)
models, which are variants of the X(5) critical point symmetry
developed within the framework of the geometric collective model,
are compared to two-parameter calculations in the framework of the
interacting boson approximation (IBA) model. The results show that
these geometric models coincide with IBA parameters consistent
with the phase/shape transition region of the IBA for boson
numbers of physical interest (close to 10). Nuclei within the
rare-earth region and select Os and Pt isotopes are identified as
good examples of X(3), X(5)-$\beta^2$, and X(5)-$\beta^4$
behavior.


Section: Nuclear Structure

\newpage

\section{Introduction} 

Critical point symmetries \cite{IacE5,IacX5}, describing nuclei at
points of shape/phase transitions between different limiting
symmetries, have recently attracted considerable attention, since
they lead to parameter independent (up to overall scale factors)
predictions which are found to be in good agreement with
experiment \cite{CZ1,ClarkE5,CZ2,ClarkX5}. The X(5) critical point
symmetry \cite{IacX5}, was developed to describe analytically the
structure of nuclei at the critical point of the transition from
vibrational [U(5)] to prolate axially symmetric [SU(3)] shapes.
The solution involves a five-dimensional infinite square well
potential in the $\beta$ collective variable and a harmonic
oscillator potential in the $\gamma$ variable. The success of the
X(5) model in describing the properties of some nuclei with
parameter free (except for scale) predictions has led to
considerable interest in such simple models to describe
transitional nuclei. Since its development, numerous extensions
involving either no free parameters or a single free parameter
have been developed. Those approaches which involve a single
parameter include replacing the infinite square well potential
with a sloped well potential~\cite{markslope}, exact decoupling of
the $\beta$ and $\gamma$ degrees of freedom~\cite{mark2005}, and
displacement of the infinite square well potential, or the
confined $\beta$-soft model~\cite{CBS}. Parameter free variants of
the X(5) model include the X(5)-$\beta^2$ and X(5)-$\beta^4$
models \cite{BonX5}, in which the infinite square well potential
is replaced by a $\beta^2$ and a $\beta^4$ potential respectively,
as well as the X(3) model \cite{X3}, in which the $\gamma$ degree
of freedom is frozen to $\gamma=0$, resulting in a
three-dimensional Hamiltonian, in which an infinite square well
potential in $\beta$ is used.

Prior to these simple geometric models, shape/phase transitions
were investigated~\cite{Gilmore} within the interacting boson
approximation (IBA) model \cite{IA} by constructing the classical
limit of the model, using the coherent state formalism
\cite{Kirson,Dieperink}. Using this method it was shown
\cite{Gilmore,Dieperink} that the shape/phase transition between
the U(5) and SU(3) limiting symmetries is of first order, while
the transition between the U(5) and O(6) ($\gamma$-unstable)
limiting symmetries is of second order. Furthermore, the region of
phase coexistence within the symmetry triangle \cite{Casten} of
the IBA has been studied \cite{IZC,Zamfir66,IZ} and its borders
have been determined \cite{Werner,Fernandes}, while a similar
structural triangle for the geometric collective model has been
constructed \cite{Zhang}.

It is certainly of interest to examine the extent to which the
parameter free (up to overall scale factors) predictions of the
various critical point symmetries and related models, built within
the geometric collective model, are related to the shape/phase
transition region of the IBA. It has already been found
\cite{McC7103} that the X(5) predictions cannot be exactly
reproduced by any point in the two-parameter space of the IBA,
while best agreement is obtained for parameters corresponding to a
point close to, but outside the shape/phase transition region of
the IBA.  In the present work we examine the extent to which the
predictions of the X(5)-$\beta^2$, X(5)-$\beta^4$, and X(3) models
can be reproduced by two-parameter IBA calculations using boson
numbers of physical interest (close to 10) and the relation of
these geometrical models to the shape/phase transition region of
the IBA. Even-even nuclei corresponding to reasonable experimental
examples of the manifestation of the X(3), X(5)-$\beta^2$, and
X(5)-$\beta^4$ models are also identified.

In Section 2 the IBA Hamiltonian used and the shape/phase
transition region within the IBA symmetry triangle are briefly
described. Predictions of the X(5)-$\beta^2$, X(5)-$\beta^4$, and
X(3) models are compared to the IBA predictions in Section 3 and
to experimental data in Section 4. Finally, Section 5 contains
discussion of the present results and implications for further
work.

\section{The IBA Hamiltonian and symmetry triangle} 

The study of shape/phase transitions in the IBA is facilitated by
writing the IBA Hamiltonian in the form \cite{Zamfir66,Werner}
\begin{equation}\label{eq:e1}
H(\zeta,\chi) = c \left[ (1-\zeta) \hat n_d -{\zeta\over 4 N_B}
\hat Q^\chi \cdot \hat Q^\chi\right],
\end{equation}
where $\hat n_d = d^\dagger \cdot \tilde d$, $\hat Q^\chi =
(s^\dagger \tilde d + d^\dagger s) +\chi (d^\dagger \tilde
d)^{(2)},$ $N_B$ is the number of valence bosons, and $c$ is a
scaling factor. The above Hamiltonian contains two parameters,
$\zeta$ and $\chi$, with the parameter $\zeta$ ranging from 0 to
1, and the parameter $\chi$ ranging from 0 to $-\sqrt{7}/2=-1.32$.
With this parameterization,  the entire symmetry triangle of the
IBA, shown in Fig.~1, can be described, along with each of the
three dynamical symmetry limits of the IBA. The parameters
$(\zeta,\chi)$ can be plotted in the symmetry triangle by
converting them into polar coordinates \cite{McC6906}
\begin{equation}\label{eq:e2}
\rho = {\sqrt{3} \zeta \over \sqrt{3} \cos\theta_\chi
-\sin\theta_\chi}, \qquad \theta = {\pi \over 3} +\theta_\chi,
\end{equation}
where $\theta_\chi=(2/\sqrt{7})\chi(\pi/3)$.

Using the coherent state formalism of the IBA
\cite{IA,Kirson,Dieperink} one can obtain the scaled total energy,
$E(\beta,\gamma)/(c N_B)$, in the form \cite{IZ}
$$ {\cal E}(\beta,\gamma)= {\beta^2 \over 1+\beta^2} \left[
(1-\zeta)-(\chi^2+1) {\zeta \over 4 N_B}\right] -{5\zeta\over 4
N_B (1+\beta^2)} $$
\begin{equation}\label{eq:e3}
-{\zeta (N_B-1) \over 4 N_B (1+\beta^2)^2} \left[ 4\beta^2 -4
\sqrt{2\over 7} \chi \beta^3 \cos 3\gamma +{2\over 7} \chi^2
\beta^4\right] ,
\end{equation}
where $\beta$ and $\gamma$ are the two classical coordinates,
related \cite{IA} to the Bohr geometrical variables \cite{Bohr}.

As a function of $\zeta$, a shape/phase coexistence region
\cite{IZC} begins when a deformed minimum, determined from the
condition ${\partial^2 {\cal E} \over \partial \beta^2}
\vert_{\beta_0\neq 0}=0$, appears in addition to the spherical
minimum, and ends when only the deformed minimum remains. The
latter is achieved when ${\cal E}(\beta,\gamma)$ becomes flat at
$\beta=0$, fulfilling the condition \cite{Werner} ${\partial^2
{\cal E} \over \partial \beta^2} \vert _{\beta=0}=0$, which is
satisfied for
\begin{equation}\label{eq:e4}
\zeta^{**}= {4 N_B \over 8 N_B +\chi^2-8}.
\end{equation}

The former, $\zeta^*$, can be derived from the results of Ref.
~\cite{catast}. For $\chi=-\sqrt{7}/2$ this point is given by

\begin{equation}
\zeta^* = {(896\sqrt{2} + 656 R)N_B \over -1144\sqrt{2} +123 R
+(1536\sqrt{2} +164R)N_B }
\end{equation}

\noindent where

\begin{equation}
R= \sqrt{ {35456\over 15129}+{32 \enskip 6^{2/3} \over 41} }
-\sqrt{ {70912\over 15129} -{32 \enskip 6^{2/3} \over 41} +
{3602816 \over 15129 \sqrt{1108+369 \enskip 6^{2/3}} } }
\end{equation}

In between there is a point, $\zeta_{\rm crit}$,  where the two
minima are equal and the first derivative of ${\cal E}_{min}$,
$\partial {\cal E}_{min}/\partial \zeta$, is discontinuous,
indicating a first-order phase transition.  For $\chi=-\sqrt{7}/2$
this point is \cite{Fernandes}
\begin{equation}\label{eq:e5}
\zeta_{\rm crit} = {16 N_B \over 34 N_B -27}.
\end{equation}

\noindent Expressions for $\zeta^*$ and $\zeta_{crit}$ involving
the parameter $\chi$ can also be deduced using the results of Ref.
~\cite{catast}.

The range of $\zeta$ corresponding to the region of shape/phase
coexistence shrinks with decreasing $\vert \chi \vert$ and
converges to a single point for $\chi=0$, which is the point of a
second-order phase transition between U(5) and O(6), located on
the U(5)--O(6) leg of the symmetry triangle (which is
characterized by $\chi=0$)  at $\zeta= N_B/(2N_B-2)$, as seen from
Eq. (\ref{eq:e4}). The phase transition region of the IBA is
included in Fig. 1.  For $N_B=10$, it is clear that the left
border of the phase transition region, defined by $\zeta^*$, and
the line defined by $\zeta_{\rm crit}$ nearly coincide. For
$\chi=-1.32$, in particular, one has $\zeta^*=0.507$ and
$\zeta_{\rm crit}=0.511$. Therefore in what follows we shall use
$\zeta_{\rm crit}$ as the approximate left border of the phase
transition region.

\section{Comparison of X(3), X(5)-$\beta^2$, and  X(5)-$\beta^4$ predictions
to the IBA} 

The most basic structural signature of the geometrical models,
X(3), X(5)-$\beta$$^2$, and  X(5)-$\beta$$^4$ is the yrast band
energy ratio $R_{4/2}$ $\equiv$ $E$($4_1^+$)/$E$($2_1^+$). Since
this is a simple and often experimentally well known observable,
we use the $R_{4/2}$ ratio as a starting point for these
calculations.  A constant value of the $R_{4/2}$ ratio can be
obtained in the IBA for a small range of $\zeta$ values (since
both provide a measure of the quadrupole deformation) and a wider
range of $\chi$ values. Figure 1 gives the loci of the parameters
which reproduce the $R_{4/2}$ ratios of X(3) (2.44),
X(5)-$\beta^2$ (2.65), X(5)-$\beta^4$ (2.77), and X(5) (2.90), for
$N_B=10$. As expected from the varying $R_{4/2}$ ratios, there is
a smooth evolution of the lines from X(3) up through X(5),
corresponding to an increase in the average $\zeta$ value.
Relating to the phase/shape transition region of the IBA, the X(3)
locus begins on the U(5)-SU(3) leg of the triangle close to the
left border of the phase/shape transition region and then crosses
it as the absolute value of $\chi$ decreases. The X(5)-$\beta^2$
locus starts within the right border of the phase/shape transition
region on the bottom leg of the triangle, then diverges slightly
away from it. The X(5)-$\beta^4$ and X(5) loci lie just beyond the
phase/shape transition region on the U(5)-SU(3) leg of the
triangle, then move away from it. This evolution can be understood
by considering the potentials used in these solutions.
X(5)-$\beta^2$ uses a harmonic oscillator potential while
X(5)-$\beta^4$ involves a potential intermediate between the
$\beta^2$ potential and the infinite square well potential of
X(5).  Note that each of these modified versions of the X(5)
solution are at some point closer to the phase/shape transition
region of the IBA than X(5) itself.

To investigate the agreement between these different models and
the IBA further, Fig. 2 illustrates some key structural
observables as a function of the parameter $\zeta$ for different
values of the parameter $\chi$. The shape/phase transition region
of the IBA, bordered by $\zeta^*$ (which almost coincides with
$\zeta_{\rm crit}$, as discussed in relation to Fig.~1) on the
left and by $\zeta^{**}$ on the right, is marked by the shaded
area. Note that the shaded area corresponds to the phase/shape
transition region for $\chi=-1.32$. As $\chi$ $\rightarrow$ 0,
this region becomes increasingly narrower. This $\chi$ dependence
is not shown in Fig. 2, since it is small.  The
parameter-independent predictions of X(3), X(5)-$\beta^2$,
X(5)-$\beta^4$, and X(5) are shown as horizontal lines. In Fig. 2
and the following discussion, the notation $E$($2_{0_2^+}^+$)
refers to the energy of $2^+$ state belonging to the $0_2^+$ band.
For the observables involving energies (left column of Fig. 2),
the X(3) and X(5)-$\beta^2$ models coincide exactly with the
predictions of the IBA in the phase transition region for $\chi$
values close to $-1.32$.  The predictions for X(3) (solid line)
intersect the IBA predictions for $\chi = -1.32$ on, or near, the
left border of the phase transition region.  For the energy ratios
given in Fig. 2, $R_{4/2}$, $E(0_2^+)$/$E(2_1^+)$, and
$E$($2_{0_2^+}^+$)/$E$($2_1^+$), the IBA predictions for
$\zeta_{crit}$ and $\chi =-1.32$ are 2.44, 2.65, and 4.02,
respectively.  These are in close agreement with the X(3)
predictions of 2.44, 2.87, and 4.83, respectively.  On the other
hand, the predictions for X(5)-$\beta^2$ (dashed line) for all
three energy ratios intersect with the IBA predictions for $\chi =
-1.32$ on, or very close to, the right border of the phase/shape
transition region.  The IBA predictions at $\zeta^{**}$ and $\chi
= -1.32$ for $R_{4/2}$, $E(0_2^+)$/$E(2_1^+)$, and
$E$($2_{0_2^+}^+$)/$E$($2_1^+$) are 2.68, 3.41, and 5.23,
respectively.  These are very similar to the X(5)-$\beta^2$
predictions of 2.65, 3.56, and 4.56, respectively. The
X(5)-$\beta^4$ and X(5) predictions for energies do not coincide
with the predictions of the IBA in the phase/shape transition
region for any set of parameter values. Overall, the intersection
of the IBA calculations with the X(5)-$\beta^4$ and X(5)
predictions lies closest to the phase transition region of the IBA
for $\chi = -1.32$ and moves further away for decreasing
$|$$\chi$$|$. In summary, for the observables involving the energy
ratios in Fig. 2, the X(3) solution is quite similar to the IBA
predictions for $\chi = -1.32$ at $\zeta_{crit}$, the
X(5)-$\beta^2$ solution corresponds closely to the IBA predictions
for $\chi=-1.32$ at $\zeta^{**}$, and the X(5)-$\beta^4$ and X(5)
solutions do not match the IBA predictions in the phase/shape
transition region for any value of $\chi$.

For the observables involving electromagnetic transition strengths
(right column of Fig. 2), the correspondence between the different
geometrical models and the IBA predictions in the phase/shape
transition region changes somewhat. For the $B$($E$2) ratios,
$B_{4/2}$ $\equiv$ $B$($E$2; $4_1^+\rightarrow2_1^+$)/$B$($E$2;
$2_1^+\rightarrow0_1^+$) and  $B_{0/2}$ $\equiv$ $B$($E$2;
$0_2^+\rightarrow2_1^+$)/$B$($E$2; $2_1^+\rightarrow0_1^+$), the
X(3) predictions do not coincide with the predictions of the IBA
in the phase/shape transition region for any value of $\chi$, with
the X(3) predictions being larger than those of the IBA in the
phase/shape coexistence region.  The X(5)-$\beta^2$ solution shows
a better correspondence with the IBA in the phase/shape transition
region for the $B_{4/2}$ and $B_{0/2}$ ratios.  The X(5)-$\beta^2$
predictions for $B_{4/2}$ and $B_{0/2}$ of 1.77 and 1.21,
respectively, are very close to the IBA predictions at
$\zeta_{crit}$ and $\chi = -1.32$ of 1.73 and 1.23, respectively.
The X(5)-$\beta^4$ solution also intersects the IBA predictions
for $B_{4/2}$ and $B_{0/2}$ for $\zeta$ values within the
phase/shape transition region.  For the final $B$($E$2) ratio,
$B$($E$2; $2_{0_2^+}^+\rightarrow0_1^+$)/$B$($E$2;
$2_{0_2^+}^+\rightarrow4_1^+$),  all three geometrical models
predict values $<$ 0.04, which is consistent with the predictions
of the IBA in within the phase/shape transition region for all
values of $\chi$.

Motivated by the above findings, we show in Fig.~3 the same
structural observables studied in Fig. ~2, but now as a function
of the IBA parameter $\chi$ and with $\zeta$ fixed to the value
$\zeta^{**}$, corresponding to the right border of the phase
transition region, for $N_B=10$.  The predictions for each of the
different geometrical models are again indicated by horizontal
lines. Each of the geometrical models intersects the IBA
predictions for $\zeta^{**}$ for one or more observables, although
the results do not coincide exactly for the complete set of
observables. For example, the X(3) predictions for the observables
$R_{4/2}$, $R_{0/2}$ $\equiv$ $E$($0_2^+$)/$E$($2_1^+$), and
$E$($2_{0_2^+}^+$)/$E$($2_1^+$) coincide exactly with the phase
transition region of the IBA for $\chi \sim -0.9$. However, this
same agreement is not obtained for the $B$($E$2) strengths.
Considering all the observables given in Fig. 3, the
X(5)-$\beta^2$ model perhaps provides the closest level of
agreement with the predictions of the phase/shape transition
region in the IBA. For a $\chi$ value of $\sim -1.2$, the energy
ratio predictions of X(5)-$\beta^2$ are very well reproduced by
the IBA calculations for $\zeta^{**}$. The $B$($E$2) ratios show
less agreement but are reproduced within an order of magnitude or
better.

In view of the above results, we compare in Fig. 4(top) the
parameter independent level scheme of X(3) to the level scheme of
the IBA with $N_B=10$, $\chi=-1.32$, and $\zeta=\zeta_{\rm
crit}=0.51$. Similarly in Fig. 4 (middle) the parameter
independent level scheme of X(5)-$\beta^2$ is compared to the
level scheme of the IBA with $N_B=10$, $\chi=-1.32$, and
$\zeta=\zeta^{**}=0.54$, and in Fig. 4 (bottom) the level scheme
of X(5)-$\beta^4$ is compared to the level scheme of the IBA with
$N_B=10$, $\chi=-1.32$, and $\zeta=0.55$.  The energy levels in
the ground state band and the excited K = $0_2^+$ band are
reproduced quite well.  Both the intra- and inter-band $B$($E$2)
strengths are consistently lower in the IBA compared with the
predictions of the geometrical models.  The decays from the $2^+$
and $4^+$ members of the K = $0_2^+$ sequence in each of the
geometrical models exhibit a pattern where spin-ascending branches
are enhanced and spin-descending branches are highly suppressed.
While the spin-descending branches are consistently highly
suppressed in the IBA predictions, the spin-ascending branches are
not as strong as given in the geometrical models.  Varying $\chi$
away from $-1.32$ can increase the strength of the spin-ascending
branches, as will be seen in the discussion of Section 4.

Till now, our focus has been on the correspondence between the
predictions of the X(3), X(5)-$\beta^2$, and X(5)-$\beta^4$ models
and the IBA predictions within the phase/shape coexistence region
and near the bottom leg of the triangle. While indeed we find
reasonable agreement between the geometrical models and the IBA
predictions within the phase/shape coexistence region, there is no
reason to constrain the analysis to a single region of the IBA
parameter space.  To investigate the level of agreement obtainable
between the geometrical models and the full two-parameter space of
IBA for $N_B$ = 10, we treat both $\zeta$ and $\chi$ as free
parameters and attempt to fit the X(3), X(5)-$\beta^2$, and
X(5)-$\beta^4$ predictions. The resulting spectra and parameters
are shown in Fig. 5.  These ``best fits'' were obtained by first
determining the range of parameters which well reproduced the
$R_{4/2}$ and $R_{0/2}$ ratios and then adjusting the parameters
within the determined range to best reproduce the $B$($E$2)
strengths and the energies of the higher spin states.  For each
model the ``best fit'' corresponds to values of $\zeta$  slightly
larger than those given in Fig. 4 and $|$$\chi$$|$ values somewhat
less than 1.32. The $R_{4/2}$ and $R_{0/2}$ ratios are reproduced
now almost exactly, since this was an initial constraint on the
fitting procedure. Comparing to the results given in Fig. 4, the
energies of the excited $0^+$ sequence are better reproduced, with
little change to the agreement for $B$($E$2) strengths.  In the
case of X(3), the IBA ``best fit'' with $\zeta = 0.55$ corresponds
exactly to the right edge of the phase/shape transition region of
the IBA ($\zeta^{**}$) for $\chi = -0.92$. This result is hinted
at in Fig. 1, where the X(3) line starts at $\zeta_{crit}$ for
$\chi = -1.32$ then moves across the phase transition region,
intersecting with $\zeta^{**}$ for $\chi \sim -0.9$. Thus, a
reasonable description of the X(3) model can be obtained for a
range of $\zeta$ and $\chi$ values within the phase/shape
transition region of the IBA.

To improve the agreement between $B$($E2$) strengths, generally a
smaller value of $\zeta$ is required in the IBA calculations. This
can be seen by again considering the right column of Fig.~2. In
the case of the X(3) model, the IBA predictions intersect the X(3)
predictions of $B$($E$2; $4_1^+\rightarrow2_1^+$)/$B$($E$2;
$2_1^+\rightarrow0_1^+$) and $B$($E$2;
$0_2^+\rightarrow2_1^+$)/$B$($E$2; $2_1^+\rightarrow0_1^+$) only
for $\chi \sim -1.32$ and $\zeta \sim 0.45$.  These parameters,
however, lead to a much more vibrational spectrum in terms of
energies ($R_{4/2}\sim 2.1$). As another example, the
X(5)-$\beta^2$ predictions for $B$($E$2;
$4_1^+\rightarrow2_1^+$)/$B$($E$2; $2_1^+\rightarrow0_1^+$) and
$B$($E$2; $0_2^+\rightarrow2_1^+$)/$B$($E$2;
$2_1^+\rightarrow0_1^+$) are well reproduced by the IBA parameters
$\chi \sim -1.0$ and $\zeta \sim 0.5$. This again leads to IBA
predictions for a vibrational spectrum in terms of energies with
$R_{4/2} \sim 2.3$.

Since we have found that the X(3),  X(5)-$\beta^2$, and
X(5)-$\beta^4$ predictions are best reproduced by IBA Hamiltonians
with $\chi=-1.32$, or close to it, it is instructive to study
\cite{Fernandes} the evolution with increasing $\zeta$ of the IBA
total energy curves for $N_B=10$ and $\chi=-1.32$, shown in Fig.
6. It is clear that at $\zeta=\zeta_{\rm crit}$, where the two
minima are equal, the total energy curve can be quite well
approximated by an infinite square well, which is the potential
used in X(3), best reproduced with $\zeta=\zeta_{\rm crit}$. In
contrast, at $\zeta=\zeta^{**}$ a deeper minimum at positive
$\beta$ starts to develop. The total energy curves at and beyond
$\zeta^{**}$ are quite similar to the $\beta^2$ and $\beta^4$
potentials, when the latter are supplemented by a
$L(L+1)/(3\beta^2)$ centrifugal term \cite{mark2005}, found
recently through the use of novel techniques allowing for the
exact numerical diagonalization of the Bohr Hamiltonian
\cite{Rowe735,Rowe45,Rowe753}. Examples of such potentials for
$L=2,6,10$ are shown in Fig.~7.

\section{Comparison of X(3), X(5)-$\beta^2$, and X(5)-$\beta^4$ predictions
to experiment} 

Several nuclei in the rare-earth region with $N=90$ have been
identified \cite{CZ2,ClarkX5} as candidates for the X(5) critical
point model. Therefore one obvious region to look for candidates
for the X(3), X(5)-$\beta^2$, and X(5)-$\beta^4$ models is in the
neighbors to these nuclei. In addition, within the framework of
the IBA, detailed fits~\cite{McC7105,McC7106} to Os and Pt
isotopes have identified nuclei which lie close to the shape/phase
transition region of the IBA. Candidates can be identified by
considering the trajectories of different isotopic chains in the
IBA symmetry triangle and the lines corresponding to the X(3),
X(5)-$\beta^2$, and X(5)-$\beta^4$ models, seen in Fig.~1, and in
addition, the best agreement for $R_{0/2}$. We identify $^{186}$Pt
and $^{172}$Os as candidates for the X(3) model, $^{146}$Ce and
$^{174}$Os as candidates for the X(5)-$\beta^2$ model, and
$^{158}$Er and $^{176}$Os as candidates for the X(5)-$\beta^4$
model. The experimental level schemes of these nuclei are compared
to the relevant geometrical model predictions as well as IBA
calculations in Figs. 8-10, while numerical values for some key
observables are given in Tables 1-3.

Considering that these nuclei were essentially chosen on the basis
of only their $R_{4/2}$ ratio and $R_{0/2}$ ratio, the level of
agreement for the other experimental observables is quite
impressive. Overall the spacings in the K = $0_2^+$ excited
sequence are well reproduced by both the geometrical models and
the IBA calculations.  The one exception is in $^{172}$Os, where
experimentally the first two levels of the K = $0_2^+$ band are
lying too close, resulting in $R_{4/2} \sim 7$, which cannot be
correct, indicating that the experimental information on these
levels should be reconsidered. The X(3) model shows excellent
agreement with the yrast band $B$($E$2) values in $^{186}$Pt and
$^{172}$Os, whereas the IBA significantly underpredicts them.
Identical results are found for the yrast band $B$($E$2) strengths
in $^{158}$Er (compared with the X(5)-$\beta^4$) model).  In
$^{174}$Os, the yrast $B$($E$2) strengths are overestimated by the
X(5)-$\beta^2$ model, while the IBA calculations provide a
reasonable description.  The branching ratios from excited states
in the K = $0_2^+$ sequence are also well reproduced by the
geometrical models.  Overall, each of these candidate nuclei
present an enhanced decay to the spin-ascending branch and a
suppression to the spin-descending branch, in agreement with the
predictions of X(3), X(5)-$\beta^2$, and X(5)-$\beta^4$.  The IBA
calculations also generally follow this pattern, with the
exception of the predictions for $^{174,176}$Os.

Given the narrowness of the shape/phase transition region of the
IBA, it is not surprising that good experimental examples of X(3),
X(5)-$\beta^2$, and X(5)-$\beta^4$ are provided by neighboring
even-even nuclei ($^{172}$Os--$^{174}$Os--$^{176}$Os). The IBA
total energy curves for these nuclei, obtained from Eq. (3) and
the parameters given in the captions of Figs. 8-10, are shown in
Fig. 11.  With increasing neutron number, the total energy curves
evolve from a shallow deformed minimum in $^{172}$Os to more
pronounced single deformed minima in $^{174,176}$Os. Qualitatively
speaking, the evolution of these potentials resembles the
evolution of the potentials one would obtain in moving from X(3)
to X(5)-$\beta^2$, to X(5)-$\beta^4$, namely a flat bottomed
potential in X(3) followed by a potential where the single
deformed minimum becomes larger, as in X(5)-$\beta^2$ and
X(5)-$\beta^4$.  More specifically, the slight preference for a
deformed minimum in $^{172}$Os suggests that it lies actually just
beyond $\zeta=\zeta_{\rm crit}$ within the IBA space. In fact, the
parameters obtained in the fit to $^{172}$Os are consistent with
the parameters obtained for the ``best fit'' of the X(3) solution
corresponding to $\zeta^{**}$.

\section{Discussion} 

In the present work the parameter independent (up to overal scale
factors) predictions of the X(5)-$\beta^2$, X(5)-$\beta^4$, and
X(3) models, which are variants of the X(5) critical point
symmetry developed within the framework of the geometric
collective model, are compared to the results of two-parameter
interacting boson approximation (IBA) model calculations, with the
aim of establishing a connection between these two approaches. The
study is focused on boson numbers of physical interest (around
10). It turns out that both X(3) and X(5)-$\beta^2$ lie close to
the U(5)--SU(3) leg of the IBA symmetry triangle and within the
narrow shape/phase transition region of the IBA. In particular,
for $\chi = -1.32$, X(3) lies close to $\zeta_{\rm crit}$, the
left border of the shaded shape/phase transition region of the
IBA, corresponding to IBA total energy curves with two equal
minima, while X(5)-$\beta^2$ lies near the right border of the
shape/phase transition region, $\zeta^{**}$, corresponding to IBA
total energy curves with a single deformed minimum. A set of
neighboring even-even nuclei exhibiting the X(3), X(5)-$\beta^2$,
and X(5)-$\beta^4$ behaviors have been identified
($^{172}$Os-$^{174}$Os-$^{176}$Os). Additional examples for X(3),
X(5)-$\beta^2$, and X(5)-$\beta^4$ are found in $^{186}$Pt,
$^{146}$Ce, and $^{158}$Er, respectively.  The level of agreement
of these parameter free, geometrical models with these candidate
nuclei is found to be similar to the predictions of the
two-parameter IBA calculations.

It is intriguing that the X(3) model, which corresponds to an {\sl
exactly separable} $\gamma$-rigid (with $\gamma=0$) solution of
the Bohr collective Hamiltonian, is found to be related to the IBA
results at $\zeta_{\rm crit}$, which corresponds to the critical
case of two degenerate minima in the IBA total energy curve,
approximated by an infinite square well potential in the model. It
is also remarkable that the X(5)-$\beta^2$ model, which uses {\sl
the same approximate separation of variables} as the X(5) critical
point symmetry, is found to correspond to the right border
($\zeta^{**}$) of the shape/phase transition region, related to
the onset of total energy curves with a single deformed minimum,
comparable in shape with the $\beta^2$ potential used in the model
in the presence of a $L(L+1)/(3\beta^2)$ centrifugal term
\cite{mark2005}.

Comparisons in the same spirit of the parameter independent
predictions of the E(5) critical point symmetry \cite{IacE5} and
related E(5)-$\beta^{2n}$ models \cite{AriasE5,BonE5}, as well as
of the related to triaxial shapes Z(5) \cite{Z5} and Z(4)
\cite{Z4} models, to IBA calculations and possible placement of
these models on the IBA-1 symmetry triangle (or the IBA-2 phase
diagram polyerdon \cite{AriasIBM2,CIPRL,CIAP}) can be illuminating
and should be pursued.

It should be noticed that the present work has been focused on
boson numbers equal or close to 10, to which many nuclei
correspond. A different but interesting question is to examine if
there is any connection between the X(3), X(5)-$\beta^2$, and/or
X(5)-$\beta^4$ models and the IBA for large boson numbers. This is
particularly interesting especially since it has been established
(initially for $N=1,000$ \cite{AriasE5}, recently corroborated for
$N=10,000$ \cite{AriasE5b}) that the IBA critical point of the
U(5)-O(6) transition for large $N$ corresponds to the
E(5)-$\beta^4$ model, i.e. to the E(5) model employing a $\beta^4$
potential in the place of the infinite well potential.

\section*{Acknowledgments}

The authors are thankful to F. Iachello and R.F. Casten for
valuable discussions. This work supported by U.S. DOE Grant No.
DE-FG02-91ER-40609.

\newpage

{\bf Figure captions}\\

\noindent {\bf Figure 1}: IBA symmetry triangle illustrating the
dynamical symmetry limits and their corresponding parameters. The
phase transition region of the IBA, bordered by $\zeta^*$ on the
left and by $\zeta^{**}$ on the right, as well as the loci of
parameters which reproduce the $R_{4/2}$ ratios of X(3) (2.44),
X(5)-$\beta^2$ (2.65), X(5)-$\beta^4$ (2.77), and X(5) (2.90) are
shown for $N_B=10$. The line defined by $\zeta_{\rm crit}$ is also
shown, lying to the right of the left border and almost coinciding
with it.
\\

\noindent {\bf Figure 2}: Evolution of some key structural
observables with the IBA parameters $\zeta$ and $\chi$ for
$N_B=10$. The predictions of X(3), X(5)-$\beta^2$, X(5)-$\beta^4$,
and X(5) are indicated in each panel by a horizontal line. The
values of $\zeta$ corresponding to the phase transition region of
the IBA, bordered by $\zeta_{\rm crit}$  (approximately equal to
$\zeta^*$)  on the left and by $\zeta^{**}$ on the right, are
marked by the shaded area. The small dependence of the phase
transition region on $\chi$
is not shown.\\

\noindent {\bf Figure 3}: The same structural observables shown in
Fig. 2 presented as functions of the IBA parameter $\chi$ for
$\zeta$ corresponding to the right border of the phase transition
region of the IBA (i.e. with $\zeta$ fixed at the $\zeta^{**}$
value) for $N_B=10$.\\

\noindent {\bf Figure 4}: Comparison of the IBA results for
$N_B=10$, $\chi=-1.32$, and $\zeta =\zeta_{\rm crit}=0.51$ with
the X(3) predictions (top). Same for X(5)-$\beta^2$, but with
$N_B=10$, $\chi=-1.32$, and $\zeta=\zeta^{**} =0.54$ (middle).
Same for X(5)-$\beta^4$, but with $N_B$ = 10, $\chi=-1.32$, and
$\zeta=0.55$ (bottom). The thicknesses of the arrows are
proportional to the respective $B$($E$2) values, which
are also labelled by their values.\\

\noindent {\bf Figure 5}:  IBA ``best fits'' to the X(3),
X(5)-$\beta^2$, and X(5)-$\beta^4$ predictions, produced by
treating $\zeta$ and $\chi$ as free parameters.  The resulting
parameters are included in the figure. The thicknesses of the
arrows are proportional to the respective $B$($E$2) values, which
are also labelled by their values.\\

\noindent {\bf Figure 6}: Evolution with $\zeta$ of IBA total
energy curves for $N_B=10$ and $\chi=-1.32$.\\

\noindent {\bf Figure 7}: $\beta^2/2$ (top) and $\beta^4/2$
(bottom) potentials, supplemented by the centrifugal potential
$L(L+1)/(3\beta^2)$,  for $L=2,6$, and 10.\\

\noindent {\bf Figure 8}: (Color online) Comparison of the
experimental data (middle) to the X(3) predictions (left) and IBA
calculations (right) for $^{186}$Pt (top) and $^{172}$Os (bottom).
The thicknesses of the arrows indicate the relative (gray arrows)
and absolute (white arrows) $B$($E$2) strengths which are also
labelled by their values. The absolute $B$($E$2) strengths are
normalized to the experimental $B$($E$2; $2_1^+\rightarrow0_1^+$)
value in each nucleus. Experimental data taken from
Refs.~\cite{186pt,186ptbe2,172os}.\\

\noindent {\bf Figure 9}: (Color online) Same as Fig.~8, but for
comparison of the experimental data (middle) to the X(5)-$\beta^2$
predictions (left) and IBA calculations (right) for $^{146}$Ce
(top) and $^{174}$Os (bottom). Experimental
data taken from Refs.~\cite{146ce,174os}.\\

\noindent {\bf Figure 10}: (Color online) Same as Fig.~9, but for
comparison of the experimental data (middle) to the X(5)-$\beta^4$
predictions (left) and IBA calculations (right) for $^{158}$Er
(top) and $^{176}$Os (bottom).
Experimental data taken from Refs.~\cite{158er,176os}.\\

\noindent {\bf Figure 11}: IBA total energy curves for $^{172}$Os
(left), $^{174}$Os (middle), and $^{176}$Os (right) obtained from
Eq. (\ref{eq:e3}) using
the parameter sets quoted in the captions of Figs. 8--10.\\

\newpage

Table 1: Comparison of experimental data for some key observables
of $^{186}$Pt and $^{172}$Os, taken from
Refs.~\cite{186pt,186ptbe2,172os}, to the X(3) predictions and IBA
calculations shown in Fig.~8.
\begin{center}
\begin{tabular}{l|ccccc}
\hline  & X(3) & $^{186}$Pt$_{exp}$ & $^{186}$Pt$_{IBA}$  &
$^{172}$Os$_{exp}$ & $^{172}$Os$_{IBA}$\\
\hline

$R_{4/2}$ & 2.44 & 2.55 & 2.45 & 2.65 & 2.59\\
$E(0_2^+)$/$E(2_1^+)$ & 2.87 & 2.46 & 3.07 & 3.32 & 3.30\\
$E(2_{0_2^+}^+)$/$E(2_1^+)$ & 4.83 & 4.16 & 5.1 & 3.55 & 4.56\\
$B_{4/2}$ & 1.90 & 1.67(20) & 1.61 & 1.50(17) & 1.60\\
\hline

\end{tabular}

\end{center}

\vspace{20 mm}

Table 2: Comparison of experimental data for some key observables
of $^{146}$Ce and $^{174}$Os, taken from Refs.~\cite{146ce,174os},
to the X(5)-$\beta^2$ predictions and IBA calculations shown in
Fig.~9.
\begin{center}
\begin{tabular}{l|ccccc}
\hline  & X(5)-$\beta^2$ & $^{146}$Ce$_{exp}$ & $^{146}$Ce$_{IBA}$
&
$^{174}$Os$_{exp}$ & $^{174}$Os$_{IBA}$\\
\hline

$R_{4/2}$ & 2.65 & 2.58 & 2.61 & 2.73 & 2.64\\
$E(0_2^+)$/$E(2_1^+)$ & 3.56 & 4.03 & 3.28 & 3.43 & 3.60\\
$E(2_{0_2^+}^+)$/$E(2_1^+)$ & 4.56 & 4.92 & 4.22 & 4.35 & 4.79\\
$B_{4/2}$ & 1.78 & - & 1.53 & 1.83(23) & 1.57\\
\hline

\end{tabular}
\end{center}

\vspace{20 mm}

Table 3: Comparison of experimental data for some key observables
of $^{158}$Er and $^{176}$Os, taken from Refs.~\cite{158er,176os},
to the X(5)-$\beta^4$ predictions and IBA calculations shown in
Fig.~10.
\begin{center}
\begin{tabular}{l|ccccc}
\hline  & X(5)-$\beta^4$ & $^{158}$Er$_{exp}$ & $^{158}$Er$_{IBA}$
&
$^{176}$Os$_{exp}$ & $^{176}$Os$_{IBA}$\\
\hline

$R_{4/2}$ & 2.77 & 2.74 & 2.68 & 2.93 & 2.83\\
$E(0_2^+)$/$E(2_1^+)$ & 4.35 & 4.20 & 4.28 & 4.45 & 4.65\\
$E(2_{0_2^+}^+)$/$E(2_1^+)$ & 5.60 & 5.15 & 6.56 & 5.50 & 6.10\\
$B_{4/2}$ & 1.69 & 1.49(7) & 1.51 & - & 1.51\\
\hline

\end{tabular}
\end{center}

\end{document}